\documentclass[reprint,nofootinbib,amsmath,amssymb,aps]{revtex4-2}
\usepackage{aas_macros}
\usepackage{graphicx}
\usepackage{dcolumn}
\usepackage{bm}
\usepackage{amsmath}
\usepackage{amssymb}
\usepackage[T1]{fontenc}
\usepackage{amsbsy}
\usepackage{color}
\usepackage{xcolor}
\usepackage{subcaption}
\usepackage{hyperref}
\usepackage[normalem]{ulem}
\usepackage{ragged2e}

\newcommand{\be}{\begin{equation}}
\newcommand{\ee}{\end{equation}}

\begin{document}

\title{Scale-dependent local primordial non-Gaussianity as a solution to the $S_8$ tension}

\author{Cl\'ement Stahl}
 \email{clement.stahl@astro.unistra.fr}
\affiliation{Universit\'e de Strasbourg, CNRS, Observatoire astronomique de Strasbourg, UMR 7550, 67000 Strasbourg, France}
\author{Benoit Famaey}
\affiliation{Universit\'e de Strasbourg, CNRS, Observatoire astronomique de Strasbourg, UMR 7550, 67000 Strasbourg, France}
\author{Oliver Hahn}
\affiliation{University of Vienna, Department of Astrophysics, Türkenschanzstraße17, 1180 Vienna, Austria}
\author{Rodrigo Ibata}
\affiliation{Universit\'e de Strasbourg, CNRS, Observatoire astronomique de Strasbourg, UMR 7550, 67000 Strasbourg, France}
\author{Nicolas Martinet}
\affiliation{Aix-Marseille Universit\'e, CNRS, Laboratoire d'Astrophysique de Marseille, UMR 7326, Marseille, France}
\author{Thomas Montandon}
\affiliation{Universit\'e de Montpellier, CNRS, Laboratoire Univers \& Particules de Montpellier (LUPM), UMR 5299, Montpellier, France}

\date{\today}

\begin{abstract}
For the last decade, several probes have pointed to a cosmological tension between the amplitude of density fluctuations extrapolated from the cosmic microwave background within the standard cosmological model and the one encapsulated by the $S_8$ parameter from large scale structure. The origin of this $S_8$ tension has not yet been elucidated and may hint at systematics in the data, unaccounted effects from baryonic physics, or new physics beyond the standard model of cosmology. Baryonic physics may in principle provide a nonlinear solution to the tension by suppressing the matter power spectrum more strongly on nonlinear scales than is traditionally assumed. Such a solution would not worsen the Hubble tension, contrary to many other proposed solutions to the $S_8$ tension. However, no realistic baryonic feedback in hydrodynamical simulations provides the needed suppression as a function of redshift. Here, we point out that a scale-dependence of local-type primordial non-Gaussianities (PNG), with significant PNG at scales of a few Mpc, can provide the needed suppression, since such PNG can suppress the power spectrum at slightly larger scales than baryons do. We demonstrate this by devising collisionless numerical simulations of structure formation in boxes of 0.5 Gpc/$h$ with scale-dependent local-type PNG. Our simple models show that, as a proof of principle, scale-dependent PNG, with a Gaussian random field for primordial density fluctuations on large scales and $f_{\rm NL} \simeq -300$ at $\lesssim 10$ Mpc scales, together with state-of-the-art baryonification of the matter power spectrum, can in principle solve the $S_8$ tension. The $S_8$ tension would then be a smoking-gun of non-trivial inflationary physics. 
\end{abstract}

\maketitle

\section{Introduction}\label{sec:Introduction}

Within the past few decades the $\Lambda$CDM (Lambda cold-dark matter) cosmological model has succeeded in explaining the majority of observations of our Universe across different scales and times, despite a few long-standing small-scale tensions \citep[e.g.,][]{2017ARA&A..55..343B,Peebles:2020bph,Famaey}. Nowadays, on cosmological scales, this model also presents some tensions between early and late time cosmological probes: namely the tension on the Hubble parameter $H_0$, of the order of $5 \sigma$ between the \textit{Planck} Cosmic Microwave Background (CMB) determination and direct measurements \citep{Riess:2016jrr,DiValentino:2021izs}, and that on the growth of structure parameter $S_8$, a combination of the matter density $\Omega_{\rm m}$ and of the amplitude of the power spectrum $\sigma_8$ (measured as the standard deviation of the amplitude of fluctuations when sampling the Universe within spheres of 8 Mpc/$h$): $S_8 =\sigma_8 \sqrt{\Omega_{\rm m}/0.3}$. The latter tension was measured in KiDS \cite{Hildebrandt:2016iqg} who found a smaller $S_8 = 0.745\pm 0.039$ than that inferred from the CMB ($S_8=0.851\pm 0.024$, \cite{Planck:2015fie}) at the $2.3\sigma$ level. The exact values have fluctuated in the
subsequent reanalyses of KiDS (e.g.~\cite{KiDS:2020suj}), DES (e.g.~\cite{Amon:2022ycy}), and Planck (e.g.,\cite{Planck:2018vyg}) but without fully alleviating the tension between weak-lensing analyses and the CMB.

Arguably, tensions and controversies have fuelled cosmology throughout its whole history \cite{philo} and the current tensions may be the hint of a new paradigm shift in our field. On the other hand, the origin of the current tensions could also stem from unaccounted systematic biases in analyses of growing complexity, or be a statistical fluctuation in yet small statistical samples for late-time probes. They may nevertheless be genuine indications of new physics beyond the standard $\Lambda$CDM model. While statistical fluctuations will be ruled out by the next generation of cosmological surveys, e.g., \textit{Euclid}, The \textit{Vera Rubin Observatory}, and \textit{Roman}, systematic biases and possible modifications to the $\Lambda$CDM model are currently under heavy scrutiny, both observationally and theoretically.

Recently Ref.~\cite{Amon:2022azi} proposed that the $S_8$ tension could arise from an incomplete description of the nonlinear physics at small scales, in the sense that current CMB and cosmic shear observations could be reconciled if the nonlinear matter power spectrum was more suppressed than classically assumed, introducing the $A_{\rm mod}$ parameter to quantify this suppression:
\begin{equation}
\label{eq:Amon}
    P(k,z)=P_{\rm L}(k,z)+A_{\rm mod}\left[P_{\rm NL}(k,z)-P_{\rm L}(k,z) \right],
\end{equation}
where $P(k,z)$ is the matter power spectrum as a function of wavenumber $k$ (in $h/{\rm Mpc}$) and redshift $z$, while $P_{\rm L}(k,z)$ and $P_{\rm NL}(k,z)$ are its linear and nonlinear contributions respectively, in the absence of feedback. This $A_{\rm mod}$ suppression then propagates to the shear correlation functions that can be computed as Hankel transforms of the convergence power spectrum which is itself related to an integral of the matter power spectrum weighted by a combination of galaxy distances (see e.g.~\cite{Kilbinger:2014cea} for a detailed review). The phenomenological approach of Eq.~\eqref{eq:Amon} encourages one to investigate small scale physical processes that could explain a dip in the power spectrum on nonlinear scales.

An obvious candidate is baryonic feedback which produces such a suppression due, e.g., to the ejection mechanisms by active galactic nuclei (AGN) which dilute matter on small scales. This hypothesis was tested in Refs.~\cite{Amon:2022azi,Preston:2023uup} who found that hydrodynamical simulations that could explain the needed suppression \citep[e.g., C-OWLS][]{Brun:2013yva} require a too large AGN feedback as compared to observational constraints \citep{McCarthy:2016mry,McCarthy:2023ism}. Interestingly, realistic baryonic feedback mostly affects scales $k$ around a few $h{/\rm Mpc}$ and is therefore unable to account for the nonlinear suppression on mildly larger scales close to $10^{-1}$ $h/{\rm Mpc}$ that would be needed to solve the $S_8$ tension. In fact, tuning the feedback amplitude to accommodate the suppression on the latter scales would result in unrealistically large suppression at smaller scales. 

A second class of candidates are scale-dependent primordial non-Gaussianities (PNG), since these mildly larger scales could actually be affected by PNG in the matter density field, therefore potentially offering another plausible explanation for the suppression of the power spectrum on mildly to very nonlinear scales, without requesting unrealistic baryonic feedback. This solution also presents the advantage of leaving the $H_0$ tension unchanged, while currently most studied extensions to $\Lambda$CDM can only reduce the $S_8$ tension at the cost of an increase in the $H_0$ tension \citep[e.g.,][]{Vagnozzi:2023nrq,Pedreira:2023qqt}. We therefore explore hereafter the possibility that the $S_8$ tension may in fact be a hint of non-trivial physics during inflation, and we thereby propose a physical interpretation of the parametric nonlinear solution of Refs.~\cite{Amon:2022azi,Preston:2023uup} that does not require unrealistic feedback.

Actually, some level of PNG is a generic prediction of {\it any} inflationary model, and this prediction is partially driving existing and upcoming cosmological surveys. PNG are often parametrized by the (local) parameter $f_{\rm NL}$ \cite{KomtsuSpergel:2001} which represents the amplitude of the quadratic correction to the Gaussian random field to describe primordial fluctuations. In the simplest single-field inflation models, PNG are expected to be of the order of only $10^{-2}$ \cite{Maldacena:2002vr}, this small value justifying the Gaussian approximation made in $\Lambda$CDM. However, many possible complex physical phenomena and additional degrees of freedom could be present in the inflationary era, and generically predict a larger $f_{\rm NL}$, that could also depend on scale \cite{LoVerde:2007ri,Wands:2010af,Byrnes:2010ft}. The non-Gaussian signals (using the idea of cosmological collider physics \cite{Arkani-Hamed:2015bza,Cabass:2024wob}) may therefore be the smoking gun of various processes potentially taking place during inflation, e.g., ultra slow roll \cite{Martin:2012pe}, features \cite{Chluba:2015bqa} or the presence of an electric field \cite{Stahl:2015gaa,Chua:2018dqh}, among many others. The precise measurements obtained from the Planck mission \cite{Planck:2019kim} and from Large Scale Structure (LSS) observations \cite{Cabass:2022ymb} have tightly constrained the Universe to be Gaussian on large scales, with bispectrum measurements broadly consistent with null values of $f_{\rm NL}$. However, it is important to remember that these constraints do not hold at smaller scales in the presence of scale-dependent PNG.

In recent years, we have studied the effect of large local-type PNG at galactic scales by devising simulations with small box-size $L=30$ Mpc/$h$ \cite{Stahl:2022did,Stahl:2023ccv,Stahl:2024jzk}, allowing us to gauge the impact of such PNG on typical galaxy-sized halo
scales and to explore whether some galactic-scale tensions could be alleviated as intuited in Ref.~\cite{Peebles:2020bph}. In particular, we have shown, with hydrodynamical simulations in such a cosmological context, that a negative $f_{\rm NL}$ at galactic scales forms simulated galaxies with more disky kinematics than in the vanilla $\Lambda$CDM case \cite{Stahl:2023ccv}, thereby potentially alleviating some small-scale tensions. Such a negative $f_{\rm NL}$ also slightly flattens the inner density profile of halos \cite{Stahl:2024jzk}. By stacking various such small-scale simulations with different initial random seeds, we also noticed that the nonlinear matter power spectrum could dwindle by $\sim 20 \%$ at nonlinear scales in the presence of a negative $f_{\rm NL}$. This motivated the present study, where we now include the large linear scales in our simulations with $L=500$ Mpc/$h$ boxes, to study whether such a suppression of the power spectrum could explain away the $S_8$ tension.

The paper is organized as follows: in Section \ref{sec:IC}, we briefly review some selected constraints on PNG at different scales, we present our phenomenological model, and we describe our implementation of the scale dependence of the PNG in the software generating initial conditions for N-body simulations (\texttt{monofonIC}\footnote{\url{https://bitbucket.org/ohahn/monofonic/}}). In Section \ref{sec:main}, we present our simulations of structure formation and our main results: the ratio of the nonlinear non-Gaussian power spectrum over the Gaussian one, superimposed with the value of $A_{\rm mod}$ required by Ref.~\cite{Preston:2023uup} to solve the $S_8$ tension. We also gauge the impact of baryons with the baryonification technique. We conclude in Section \ref{sec:ccl}, and also briefly discuss consequences of such models on the halo mass function and void statistics.

\section{Scale-dependent primordial non-Gaussianities}\label{sec:IC}

In a universe that is statistically homogeneous and isotropic on large scales, the small primordial fluctuations are depicted as a random field for the contrast density $\delta(\mathbf{x}) = (\rho(\mathbf{x})-\rho_0)/\rho_0$,  where $\rho_0$ represents the background density. This random field is most effectively described in Fourier space, hence with the Fourier transform of the contrast density, noted $\delta(\mathbf{k})$. The variance of $|\delta(\mathbf{k})|$ corresponds to the Fourier transform of the two-point correlation function, i.e. to the power spectrum $P(k) \propto \langle |\delta(\mathbf{k})|^2 \rangle_{|\mathbf{k}|=k}$. If the random field is Gaussian, the entirety of the statistical information is encapsulated within its two-point correlation function and hence in its power spectrum, whilst all odd correlation functions are zero. This same principle extends to the gravitational potential engendered by the contrast density field, such that all information on a Gaussian random field $\Phi_G$ for the Newtonian gravitational potential is contained within its two-point correlation function:
\begin{equation}
 \langle \Phi_{\rm G}(\mathbf{k}_1) \Phi_{\rm G}(\mathbf{k}_2) \rangle = (2\pi)^3 \delta_D(\mathbf{k}_1+\mathbf{k}_2) P(k_1) \, , \label{eq:2ptcorr}
\end{equation}
where $\delta_D$ denotes the Dirac delta function and $ \langle \dotsb \rangle$ is the ensemble average often traded for a spatial average owing to an ergodicity hypothesis. If the random field is non-Gaussian, however, the three-point correlation function and its corresponding bispectrum can display non-zero values, especially if the random field is skewed.
It is the small values measured from Planck \cite{Planck:2019kim} and LSS \cite{Cabass:2022ymb} that allow to ascertain that the Universe is close to Gaussian on large scales, but as we shall briefly review below, constraints are much less stringent on small scales.

In the standard cosmological model, the primordial power spectrum for the Newtonian potential adopts the form $P_\Phi \propto A_S {k}^{n_S-4}$, where the parameters $A_S$ and $n_S$ quantify, respectively, the amplitude of the primordial perturbations and the deviation from scale invariance ($n_S=1$ for scale invariance). Thanks to the translation invariance of the linearized equation of motion, modes are decoupled and one can linearly propagate these primordial perturbations
as $\delta(\mathbf k,t) = T(k,t) \Phi(\mathbf k) \,$, where $t$ is the cosmic time and $T(k,t)$ is the transfer function. However, when entering the nonlinear regime, one then needs to follow the evolution of perturbations through numerical simulations, as we will do in this paper.

PNG can be simply expressed as deviations from Gaussianity through a sum of the Gaussian $(\Phi_{\rm G}$) and non-Gaussian ($\Phi_{\rm NG}$) contributions:
\begin{equation}
\label{eq:PNGtemplate}
 \Phi(\mathbf x)=\Phi_{\rm G}(\mathbf x) + \Phi_{\rm NG}(\mathbf x)\,.
\end{equation}
Local templates are usually defined as a perturbative expansion around the Gaussian term:
\begin{equation}
\label{eq:localPNGtemplate}
 \Phi_{\rm NG}(\mathbf x) = - f_\mathrm{NL} \left( \Phi_{\rm G}^2(\mathbf x)  - \langle \Phi_{\rm G}^2 \rangle \right) \,.
\end{equation}
where the $f_{\rm NL}$ parameter quantifies the amplitude of the quadratic correction. Note that this expansion is done here around the Gaussian random field for the Newtonian potential $\Phi$, opposite to the Bardeen potential. The minus sign in Eq.~\eqref{eq:PNGtemplate} thus implies that a large positive value of $f_{\rm NL}$ corresponds to a distribution of the primordial contrast density skewed towards overdensities, whilst a large negative value corresponds to a contrast density field skewed towards underdensities, exactly as per the standard convention \cite{KomtsuSpergel:2001}. Most often, motivated by inflationary slow roll conditions, constraints on this parameter are set by assuming it to be constant over all scales. This $f_{\rm NL}$ parameter can also be defined as the ratio of the bispectrum to the power spectrum, and in more general local models, it can actually depend on scale \cite{Wands:2010af}.
Hereafter, we are interested in such a running parameter $f_{\rm NL}(k)$ that grows with $k$, and hence becomes significant only at small scales.

\begin{figure}
 \centering
\includegraphics[width=0.5\textwidth]{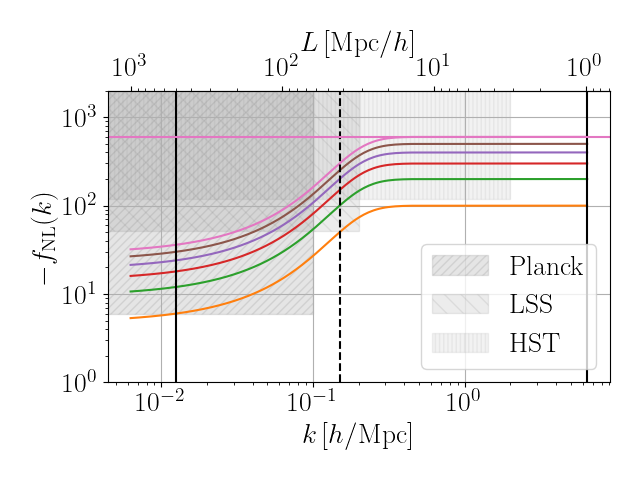}
     \caption{\justifying The values of $f_{\rm NL}(k)$ for the main models studied in this work. The white region is completely safe from all constraints while the grey zone is above the 1$\sigma$ deviation from the constraints from Planck \cite{Planck:2019kim}, LSS \cite{Cabass:2022ymb}, and HST \cite{Sabti:2020ser} . The solid lines represent 6 different values of $f_{\rm NL}^0 \in [-600\,; -100]$ for a rather smooth transition with $\sigma=0.1\,h/$Mpc at $k_{\rm min}=0.15\,h/$Mpc in Eq.~\ref{eq:fNLk} with $\alpha=1$. Each model asymptotes to the value $f_{\rm NL}^0$ at large $k$. The solid vertical lines represent the range of $k$ simulated in this work and the dotted vertical line represents the transition $k_{\rm min}=0.15\,h/$Mpc.}
     \label{fig:constraints}
 \end{figure}

\paragraph*{\bf Constraints on PNG at small scales} The parameter $f_{\rm NL}$ is tightly constrained from large scale observation of the CMB and LSS, which we illustrate in Fig.~\ref{fig:constraints}. At smaller scales, $\mathcal{O}(10)$ Mpc, the constraints get looser. Using spectral $\mu$-distortions from the CMB photons and correlating them with temperature fluctuations, $f_{\rm NL} < 6800$ at 2$\sigma$ for scales $\mathcal{O}(1)$ kpc \cite{Bianchini:2022dqh}. Using data from the Hubble Space Telescope (HST), Ref.~\cite{Sabti:2020ser} finds $f_{\rm NL} = 71^{+426}_{-237}$ at 2$\sigma$ \cite{Sabti:2020ser} with scale-dependent PNG featuring a Heaviside function activated at scales smaller than 60 Mpc. We illustrate this constraint on Fig.~\ref{fig:constraints} too. However note that when PNG are only present at scales smaller than 6 Mpc, Ref.~\cite{Sabti:2020ser} actually finds a best-fit $f_{\rm NL} \simeq -1000$, with a departure from $f_{\rm NL} = 0$ significant at 1.7$\sigma$. 

\paragraph*{\bf Phenomenological modeling}
For simplicity, we keep a local for the non-Gaussian template and assume that $f_{\rm NL}$ depends only on scale (and not on shape) such that, in Fourier space, the non-Gaussian part now reads

\begin{equation}\label{eq:NGk}
    \Phi_{\rm NG}(\mathbf k)\!\!=\!\!-f_{\rm NL}(k)\!\!\!\int\!\! \frac{d^3k_1d^3k_2}{(2\pi)^3} \delta_D(\boldsymbol k - \boldsymbol k_1 - \boldsymbol k_2)  \Phi_{\rm G}(\boldsymbol k_1) \Phi_{\rm G}(\boldsymbol k_2).
\end{equation}

In order to model the scale dependence, we choose a simple \textit{effective} three-parameter function:
\begin{equation}
\label{eq:fNLk}
f_{\rm NL}(k)=\frac{f_{\rm NL}^0}{1+\alpha}\left[\alpha+\tanh \left(\frac{k-k_{\rm min}}{\sigma} \right) \right],
 \end{equation}
where $f_{\rm NL}^0$ is the amplitude of the non-Gaussian signal on small scales (large $k$). Here, we choose $\alpha=1$ for simplicity. This choice leaves some amount of non-Gaussianity on large-scales, at the limit of current constraints, but it would of course also be possible to choose $\alpha(k_{\rm min},\sigma)$ such as to cancel $f_{\rm NL}$ completely for small $k$. The parameter $\sigma$ controls the sharpness of the transition ($\sigma \rightarrow 0$ recovers the Heaviside case \cite{Sabti:2020ser}). $k_{\rm min}$ is the typical scale of the transition between small and large PNG. Several possible non-trivial inflation mechanisms do lead to a strong scale-dependence in $f_{\rm NL}$: a time-dependent speed of sound \cite{Khoury:2008wj}, a sharp transition between a massive and a massless spectator scalar field \cite{Riotto:2010nh}, changing the initial state of the inflaton \cite{Chen:2006nt}, anisotropic inflation \cite{Dey:2013tfa}, curvatons self-interactions \cite{Byrnes:2011gh}, tachyonic instabilities \cite{McCulloch:2024hiz}, Dirac-Born-Infeld terms \cite{Chen:2005fe} arising from brane inflation, where in that case, the scale dependence of the PNG gives information about the geometry of the extra dimensions \cite{LoVerde:2007ri}. We leave for future work the exploration of the links between the phenomenological parameters $(f_{\rm NL}^0, \sigma, k_{\rm min},\alpha)$ and parameters related to the microphysics of inflation.

 In Figure \ref{fig:constraints}, we display the main constraints discussed previously along with the shapes of $f_{\rm NL}(k)$ simulated in this work. They correspond to negative $f_{\rm NL}^0 \in [-600\,; 0]$. The three other parameters are fixed: $\sigma=0.1\,h/$Mpc, $\alpha=1$ and $k_{\rm min}=0.15\,h/$Mpc. As the constraints from the different probes span a wide range of scales, we argue that those models are not \textit{a priori} ruled out in the sense that each $k$-bin used to put the constraints carries different signal to noise, and $f_{\rm NL}$ usually comes from the larger $k$-bins. Remember that Ref.~\cite{Sabti:2020ser} actually found $f_{\rm NL} \simeq -1000$ when PNG are only present at the smallest scales. A detailed analysis of the models template fitted with CMB, LSS and HST data is beyond the scope of this article, which presents the qualitative picture. A full exploration of the parameter space of scale-dependent non-Gaussian templates fitted to observations will be the topic of future explorations.

\paragraph*{\bf Numerical configuration}
Our implementation in \texttt{monofonIC} \cite{Hahn:2011uy,Michaux:2020yis} roughly follows the procedure described in Ref.~\cite{Stahl:2022did}, where instead of Eq.~\eqref{eq:localPNGtemplate}, we now use Eq.~\eqref{eq:NGk} and Eq.~\eqref{eq:fNLk}. 
Note that we have also implemented power laws: $f_{\rm NL}(k)=f_{\rm NL}^0 \times k^{n_{f_{\rm NL}}}$ in a branch of \texttt{monofonIC} but its application to the $S_8$ tension was less straightforward. The results were indeed not converged when varying the box size. This was due to non-perturbative matter fluctuations at small scales. This problem is nicely solved when using templates featuring a plateau at small scales such as Eq.~\eqref{eq:fNLk}. Another alternative to the plateau would be to set back $f_{\rm NL}(k)$ to zero at small scales (very large $k$) in order to have a bump-like feature that could possibly connect more easily to inflationary physics. If doing so at very small scales, it would not change our present results.

We compute the displacement field using Lagrangian perturbation theory at second order (2LPT) at redshift $z=32$ on a $512^3$ grid. For all our simulations, we use the following cosmological parameters compatible with Planck cosmology: $\Omega_{\rm m}=0.31$, $\Omega_{\Lambda}=0.69$, $H_0=67.7$ $\rm{km}\,\rm{s}^{-1}\,\rm{Mpc}^{-1}$, $A_S=2.11 \times 10^{-9}$ and $n_S=0.967$. The parameters describing the PNG are $f_{\rm NL}^0 \in [-600\,; 0]$, $\sigma=0.1\,h/$Mpc, $\alpha=1$ and $k_{\rm min}=0.15\,h/$Mpc. At $z=32$, the typical suppression of the power spectrum at large $k$ is always less than 2\%.

Our box length is $L=500$ Mpc/$h$, though we have checked that our main result (in Fig.~\ref{fig:main}) does not depend on the box size by re-simulating, with the same number of particles and the same random seed, boxes of length 100, 200 and 1000 Mpc/$h$. We have also checked that the choice of random seed and the fixing/pairing \cite{Angulo:2016hjd} procedure impact our main result (in Fig.~\ref{fig:main}) by less than 1\%. We have finally verified that our choice of initial redshift, even in the case of large small-scale PNG, does not lead to significant transients at the level of the matter power spectrum at $z=0$. We have run extra simulations starting at $z=200, 100, 50$ and 20: in the non-Gaussian case, we find that at redshifts 200 and 100, large transients lead to numerical inaccuracy, in particular, the large scale behavior of the matter power spectrum is off by up to 5\%, while the small scales are less affected. On the other hand, for initial redshifts 50 and 20, the $z=0$ power spectra agree below the percent level at all scales.

\section{Results}\label{sec:main}

 We evolve our initial perturbations using the Tree-PM code \texttt{Gadget 4} \cite{Springel:2020plp} with a gravitational softening length of $\epsilon=50$ $h^{-1}$kpc. We then compute the nonlinear matter power spectrum at $z=0.25$ and $z=0$. At the smallest scales we are considering in this work, baryons are however known to impact the formation of structures, the dominant physical process being the ejection of gas from the AGN. In Refs.~\cite{Amon:2022azi,Preston:2023uup}, the authors explored whether the physical process driving the suppression of the matter power spectrum could be the baryonic effects. They concluded that the effect of baryons, when
calibrated on observed quantities, is generically too weak to account for such a drop in the matter power spectrum. Even if larger feedback became allowed by observations, matching feedback on the scale of interest for PNG would dramatically oversurpress the smallest scales power spectrum.

 \begin{figure}
     \centering
     \includegraphics[width=0.5\textwidth]{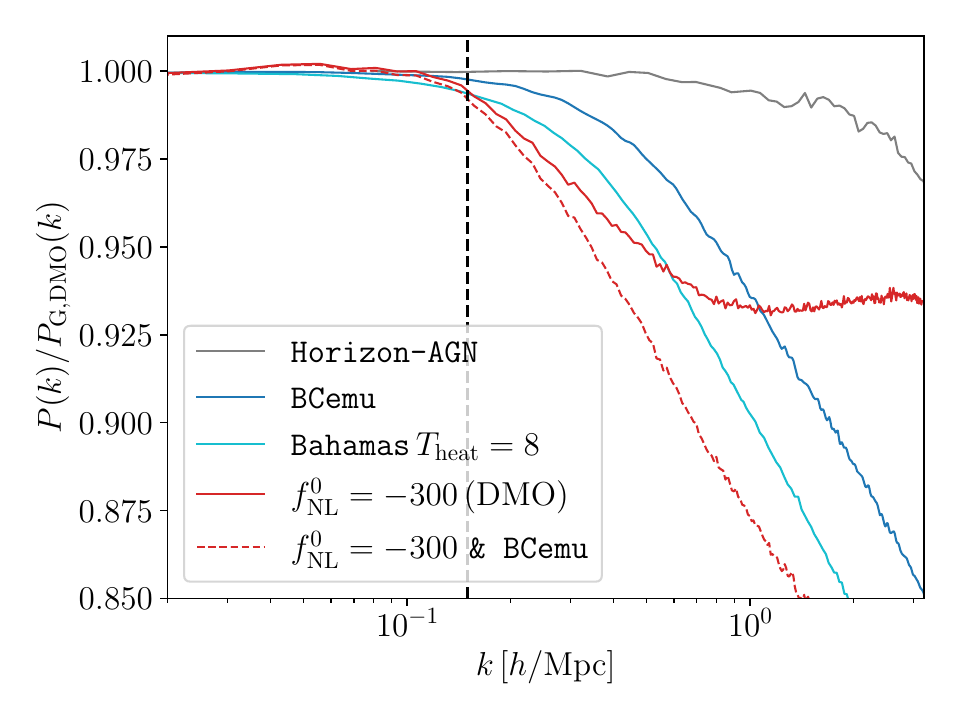}
     \caption{\justifying At $z=0$, power suppression due to baryonic feedback in different hydrodynamical simulations and the emulator \texttt{BCemu}, along with the suppressions in the scale-dependent PNG simulation with $f_{\rm NL}^0=-300$ studied in this work: Dark Matter Only (DMO) and with emulated baryonic physics. The dashed
vertical line corresponds to $k_{\rm min}=0.15\,h/{\rm Mpc}$.}
     \label{fig:simusHydro}
 \end{figure}

In Fig.~\ref{fig:simusHydro}, we illustrate this by showing the suppression of the power spectrum at $z=0$ that we get from PNG alone with $f_{\rm NL}^0 = -300$ (without taking into account any additional effect of the baryons) with respect to the Gaussian case, which we compare to the suppression related to feedback in two state-of-the-art simulations reproducing most small-scale observations \cite{Dubois:2014lxa,McCarthy:2016mry}. It is immediately clear that our PNG template can suppress the power spectrum on slightly larger scales than baryonic feedback. Note that for simulations with scale-independent $f_{\rm NL}$, such a drop of the power spectrum was already witnessed \cite{Dalal:2007cu, Smith:2010fh}, but the large scale behavior of such templates is tightly constrained. 
 
To add the baryonic effects on top of the PNG, we use an emulator. Various such emulators have been developed \cite{Mead:2020vgs,Euclid:2020rfv,Arico:2020lhq,Sharma:2024kwj}.
Baryonification \cite{Schneider:2015wta} is a technique that mimics the effect of the baryons in dark-matter-only simulation by slightly displacing the dark matter particles. In order to have a flavor of the impact of the baryonic effects on our PNG models, we use the emulator \texttt{BCemu}\footnote{\url{https://github.com/sambit-giri/BCemu}} \cite{Schneider:2018pfw,Giri:2021qin}. We use the 7 vanilla values of the parameters governing the baryonic physics: five parameters related to the gas: $M_c= 10^{13.3}$, $\mu=0.93$, $\theta_{\rm ej}=4.2$, $\gamma=2.25$, $\delta=6.4$, and two parameters related to the stars: $\eta=0.15$ and $\eta_{\delta}=0.14$, see Table 1 of Ref.~\cite{Schneider:2018pfw} for more details. A generic feature of the baryonic effects on the power spectrum is that they nicely decouple from the cosmology except from the value of the cosmological baryon fraction. While these results would need to be confirmed with hydrodynamical simulations \citep[see][]{Stahl:2023ccv} in large boxes including scale-dependent PNG, we will assume that the two effects add linearly for the present work, and leave to further investigations a confirmation with hydrodynamical simulations\footnote{Ref.~\cite{Elbers:2024dad} recently discussed this issue in the context of decaying-dark matter and massive neutrinos}. In Fig.~\ref{fig:simusHydro}, it is then clear that the PNG+baryon suppression with $f_{\rm NL}^0 = -300$ is more important and kicks in at larger scales (smaller $k$) than with realistic baryonic feedback alone.

\begin{figure*}
     \centering
     \includegraphics[width=\textwidth]{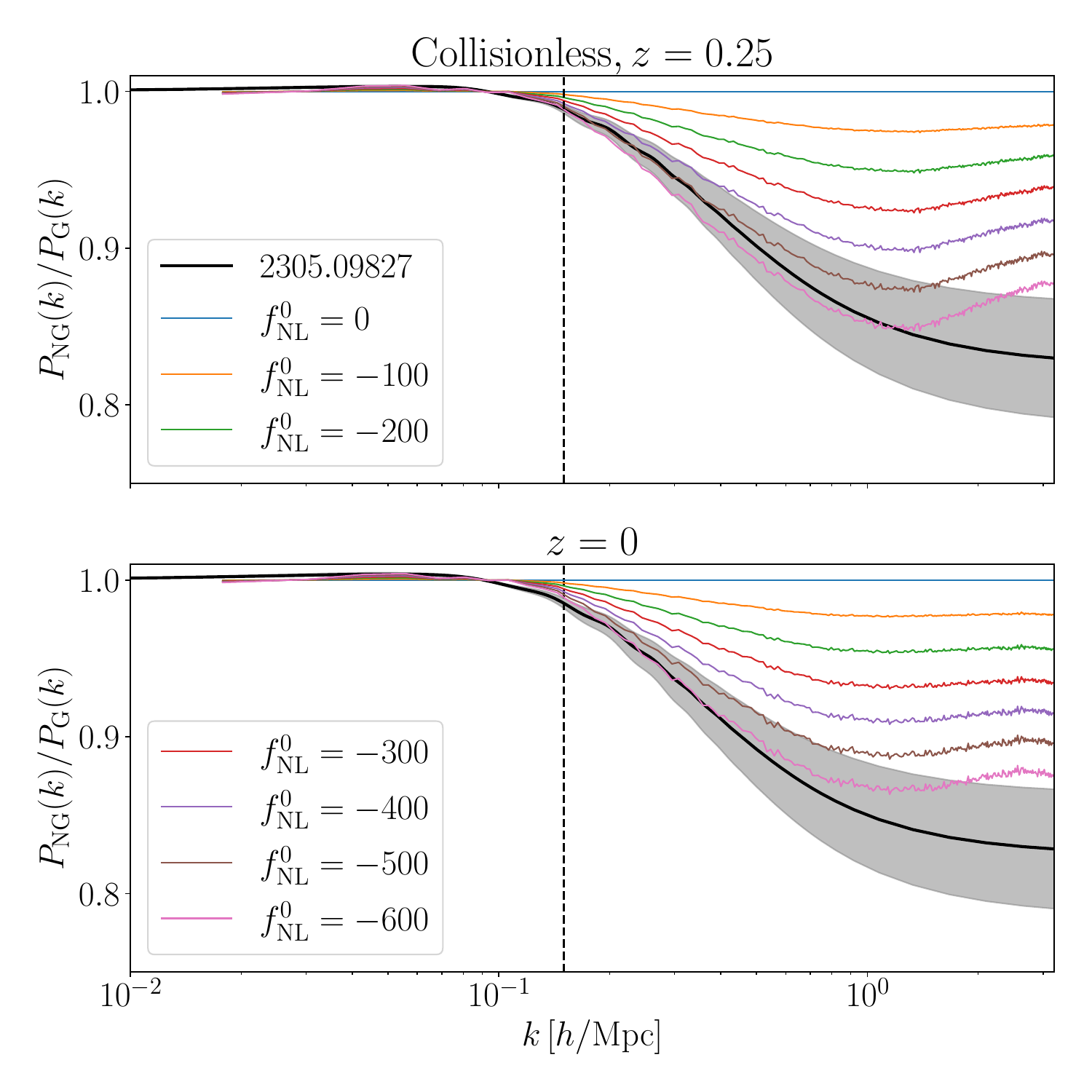}
     \caption{\justifying Power suppression in the scale-dependent PNG simulations (thin colored lines) compared to the suppression inferred from KiDS and DES to solve the $S_8$ tension (thick black line). No baryonification has been considered, i.e., with a (non-)effect of baryons comparable to, e.g., Horizon-AGN. The maximal value of $k$ represents the Nyquist wavenumber and the dashed
vertical line corresponds to $k_{\rm min}=0.15\,h/{\rm Mpc}$. Depending on the weight of the redshift-bins in the data, solving  the $S_8$ tension may require $f_{\rm NL}^0 \sim -600$ or $-500$.}
     \label{fig:main}
 \end{figure*}
 
 In Figure \ref{fig:main}, we now display the impact of our non-Gaussian models on the nonlinear matter power spectrum compared to the needed suppression to explain away the $S_8$ tension. Using \texttt{CLASS} \cite{Blas:2011rf} and \texttt{HaloFit} for the linear and nonlinear power spectra, we reproduce Eq.~(6) of Ref.~\cite{Preston:2023uup} with $A_{\rm mod}=0.82 \pm 0.04$. This value of $A_{\rm mod}$ is valid for redshift around $z \sim 0.25$ where most of the statistical power of DES weak lensing data is present \cite{Preston:2023uup}. In this Figure, no baryonification is applied to the simulations (or, equivalently, the baryon impact is negligible as in the Horizon-AGN simulation \cite{Dubois:2014lxa}, see also Ref.~\cite{Terasawa:2024agq}), in this case $f_{\rm NL}^0 \sim -600$ or $-500$ can produce the necessary drop in amplitude for the power spectrum at nonlinear scales, while leaving the linear power spectrum untouched.

\begin{figure*}
     \centering
     \includegraphics[width=\textwidth]{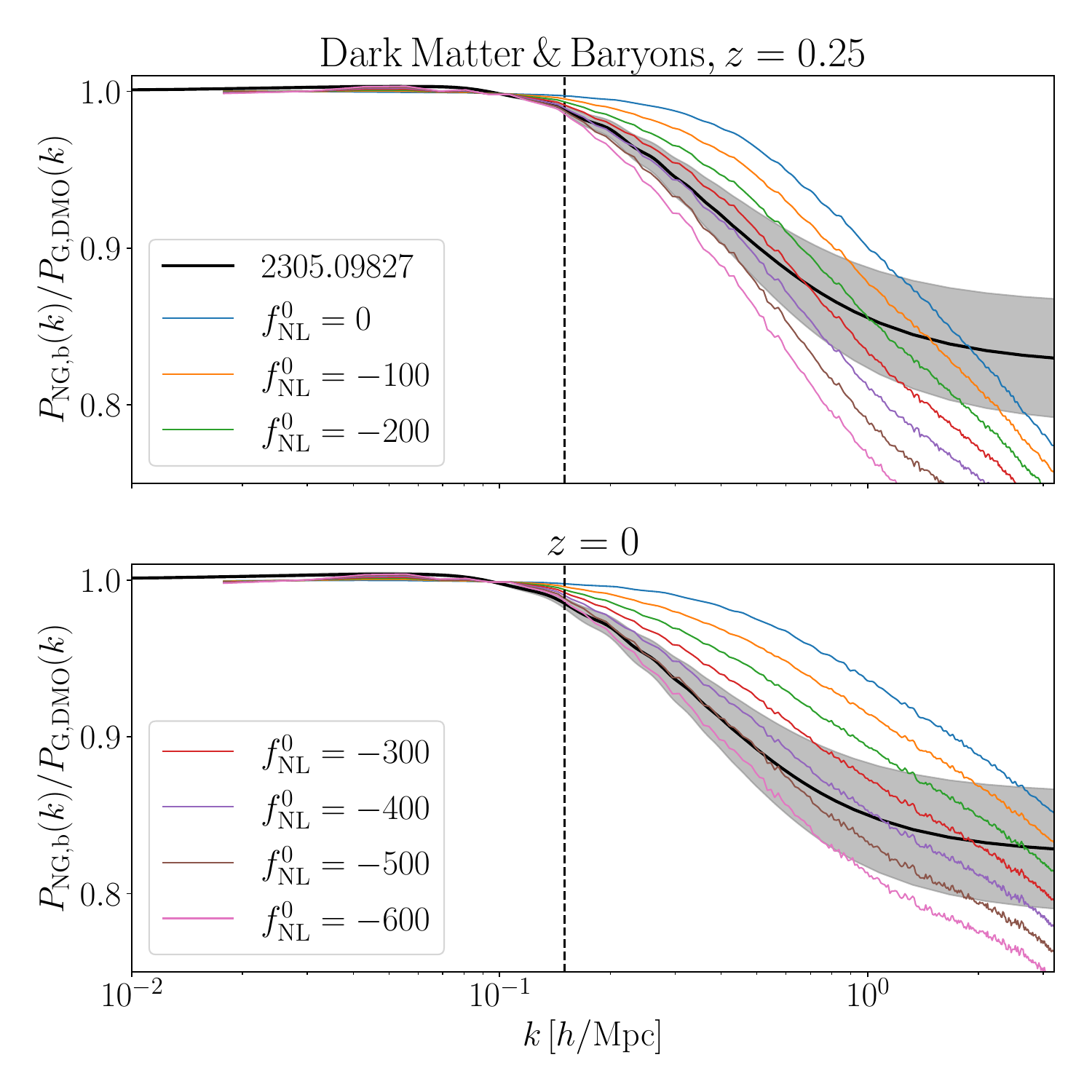}
     \caption{\justifying Same as Fig~\ref{fig:main}, but this time with the baryonification model of Ref.~\cite{Giri:2021qin}. Depending on the weight of the redshift-bins in the data, the preferred value to solve the $S_8$ tension would now be $f_{\rm NL}^0 \sim -300$ or $-400$.}
     \label{fig:mainB}
 \end{figure*}

In Figure \ref{fig:mainB}, we then display the impact of our non-Gaussian models on the nonlinear matter power spectrum along with the emulated baryonic effects from \texttt{BCemu}. In this case, depending on the weight of the redshift-bins in the data, the preferred value to solve the $S_8$ tension would be $f_{\rm NL}^0 \sim -300$ or $-400$.

To explore how much our conclusions depend on the specific models considered, we then varied the value of $k_{\rm min}$ and found (logically) that it is proportional to the value at which the nonlinear suppression of the matter power spectrum occurs. We have also varied the value of $\sigma$. In Appendix \ref{sec:appendix}, we provide a figure to illustrate the impact of the parameters.

 \section{Discussion and conclusions}
 \label{sec:ccl}

Since no realistic baryonic feedback in hydrodynamical simulations provides a suppression of the matter power spectrum at large enough scales to solve the $S_8$ tension, we explored whether a scale-dependence of local-type PNG, with negative $f_{\rm NL}$ of the order of a few $-10^2$ at $\lesssim 10$~Mpc scales, can provide the needed suppression. For this, we devised collisionless numerical simulations of structure formation in boxes of 0.5~$h^{-1}$Gpc with a simple effective template for the scale-dependence of $f_{\rm NL}(k)$. We thereby showed that such scale-dependent PNG can indeed provide the needed suppression. With a negligible effect of baryons on the matter power spectrum, the value of $f_{\rm NL}$ at small scales should be $\sim -600$ or $-500$. With state-of-the-art baryonification, a value $\sim-300$ or $-400$ would be preferred. 

The simple effective templates studied in this work provide a qualitative proof-of-concept, but of course it will need to be followed by a more complete exploration of scale-dependent non-Gaussian templates directly fitted to observations at different scales. For this, it would be convenient to devise an emulator of the effect of the scale-dependent PNG, coupled with an emulator of baryonic physics. This procedure will need to be backed by large hydrodynamical simulations \citep[see][for such hydrodynamical simulations on small scales]{Stahl:2023ccv}, along the lines of Ref.~\cite{Elbers:2024dad}. This would then allow to infer directly from weak lensing data the preferred values of the cosmological parameters along with the value of $f_{\rm NL}^0$ and possibly $k_{\rm min}$ and $\alpha$. 

In this work, we have mainly concentrated on the $S_8$ tension as deduced from weak-lensing analyses, which are less dependent on biases -- such as the infamous hydrostatic bias -- in cluster count analyses. However, the tail of the Halo Mass Function (HMF) is in principle also altered by the models studied in this work. This is also supported by early weak-lensing peak statistics analyses who found a high sensitivity of the high-mass peaks to $f_{\rm NL}$ \citep{Marian:2010mh, Hilbert:2012gr}. In Appendix \ref{app:HMF} we present, along with the Gaussian HMF, the HMF for $f_{\rm NL}^0=-300$ that shows a drop at high masses which could also impact the $S_8$ tension as studied from cluster counts \cite{Douspis:2018xlj,Salvati:2019zdp,Ghirardini:2024yni}. We also checked for the abundance of voids, since our PNG template initially favours under-densities, and find that the largest ($\sim 20$ Mpc$/h$ in our $\sim 500$ Mpc$/h$ box) voids in the $f_{\rm NL}^0=-300$ simulation are more numerous at redshift zero than in the Gaussian simulation. In addition, we have also checked that $f_{\rm NL}^0 \sim -300$ leads to a drop of the power spectrum at $z=3$ approximately compatible with the recent values for ($\Delta_{\rm lin}^2$, $n_{\rm lin}$) found from Lyman-alpha data \cite{Rogers:2023upm}. Finally, it should be noted that measurements of gravitational lensing of the CMB with ACT \cite{ACT:2023dou} and SPT-3G \cite{SPT:2023jql} have yielded a higher $S_8$ than weak-lensing analyses from LSS. These studies are however not sensible to the same redshifts. Moreover, while the ACT baseline analysis does incorporate the modeling of non-linear scales, their constraints that use only linear theory yield consistent results \cite{ACT:2023dou}, thereby showing that such analyses are, in fact, mainly
sensitive to linear scales where our PNG templates would also give a $S_8$ parameter consistent with vanilla $\Lambda$CDM.

Local PNG impact the clustering of cosmological tracers, implying a scale-dependent bias \cite{Dalal:2007cu}. It leads to tight constraints on PNG eg.~\cite{eBOSS:2021jbt}. While our models screen the large scale so that the CMB constraints lose support, a measure of the scale-dependent bias could also constrain the models studied in this work. It has been beyond the scope of the present paper to provide a complete analysis of the scale-dependent bias, which would need a larger effective volume to study the effect \cite{Avila:2022pzw,Coulton:2022qbc,Adame:2023nsx,Anbajagane:2023wif,Hadzhiyska:2024kmt}, whilst on the largest scales, the relativistic effects may also alter the measured signal \cite{Montandon:2024mku}. It will however be particularly interesting to test this in the future.

We conclude that the $S_8$ tension might, in fact, be a smoking-gun of non-trivial inflationary physics leading to scale-dependent PNG. This proposed solution to the $S_8$ tension is attractive because extensions to $\Lambda$CDM involving modified gravity rather typically lead to a boost and not to a drop of the matter power spectrum \cite{Fiorini:2023fjl, Saez-Casares:2023olw,Sletmoen:2024are}. As the present solution to the $S_8$ tension leaves the largest scales mostly untouched, it is easier to combine it with other solutions to cosmic tensions such as early dark energy solutions to the $H_0$ tension \cite{Poulin:2018cxd}, which should be investigated in the future. Such a combination has never been tried and might prove particularly promising \citep{Vagnozzi:2023nrq}.

\acknowledgments
BF, RI, and CS acknowledge funding from the European Research Council (ERC) under the European Union's Horizon 2020 research and innovation program (grant agreement No.\ 834148). TM is supported by funding from the European Research Council (ERC) under the European Union’s HORIZON-ERC-2022 (grant agreement no. 101076865). TM and CS are grateful for insightful discussions on tensions in cosmology with Vivian Poulin. CS and BF thank Théo Bruant and Fabien Castillo for stimulating exchanges. This work has made use of the Infinity Cluster hosted by the Institut d'Astrophysique de Paris.

\subsection*{Softwares}
The analysis was partially made using  \href{https://yt-project.org/}{YT} \cite{Turk:2010ah}, as well as IPython \cite{Perez:2007emg}, Matplotlib \cite{Hunter:2007ouj}, NumPy \cite{vanderWalt:2011bqk}. All power spectra presented in this articles are measured with \texttt{Pylians} \cite{Pylians}.

\subsection*{Authors' Contribution}
CS implemented in \texttt{monofonIC}
the non-Gaussian initial conditions helped by TM. The simulations presented in this work
were performed and analysed by CS in consultation with BF. CS and BF drafted the manuscript with inputs from NM. All the authors improved it by their insightful comments.

\subsection*{Carbon Footprint}
Based on the methodology of Ref.~\cite{berthoud} to convert\footnote{including the global utilisation of the cluster and the pollution due to the electrical source, the conversion factor is 4.7 gCO2e/(h.core)} the CPU hours used for the simulations presented in this work, we estimate that we have emitted 0.5 TCO2eq.

\appendix
\section{Impact of the phenomenological modeling}
\label{sec:appendix}
In our fiducial models, we fixed the phenomenological parameters of Eq.~\eqref{eq:fNLk} to $\sigma=0.1\,h/{\rm Mpc}$ and $k_{\rm min}=0.15\,h/{\rm Mpc}$. Here, we study the impact of varying those two parameters by performing more simulations as illustrative examples, still keeping $\alpha=1$ and fixing $f_{\rm NL}^0=-500$. Our results are displayed in Figure \ref{fig:impactPM}. For a sharp transition ($\sigma=0.02\,h/$Mpc, close to Heaviside), we obtain the suppression in the power spectrum for $k>k_{\rm min}$ but also a small (5\%) bump in the power spectrum around $k \sim 0.1\, h$/Mpc. We interpret it as a numerical artefact due to the $k$-binning in \texttt{monofonIC} and for this reason we use in our fiducial models a rather smooth transition such as $\sigma=0.1\,h$/Mpc. The scale $k_{\rm min}$ controls the scale at which the suppression due to PNG occurs and it is also a relevant parameter for building a robust candidate to solve the $S_8$ tension.

\counterwithin{figure}{section}
\setcounter{figure}{0}
  \begin{figure*}
     \centering
     \includegraphics[width=\textwidth]{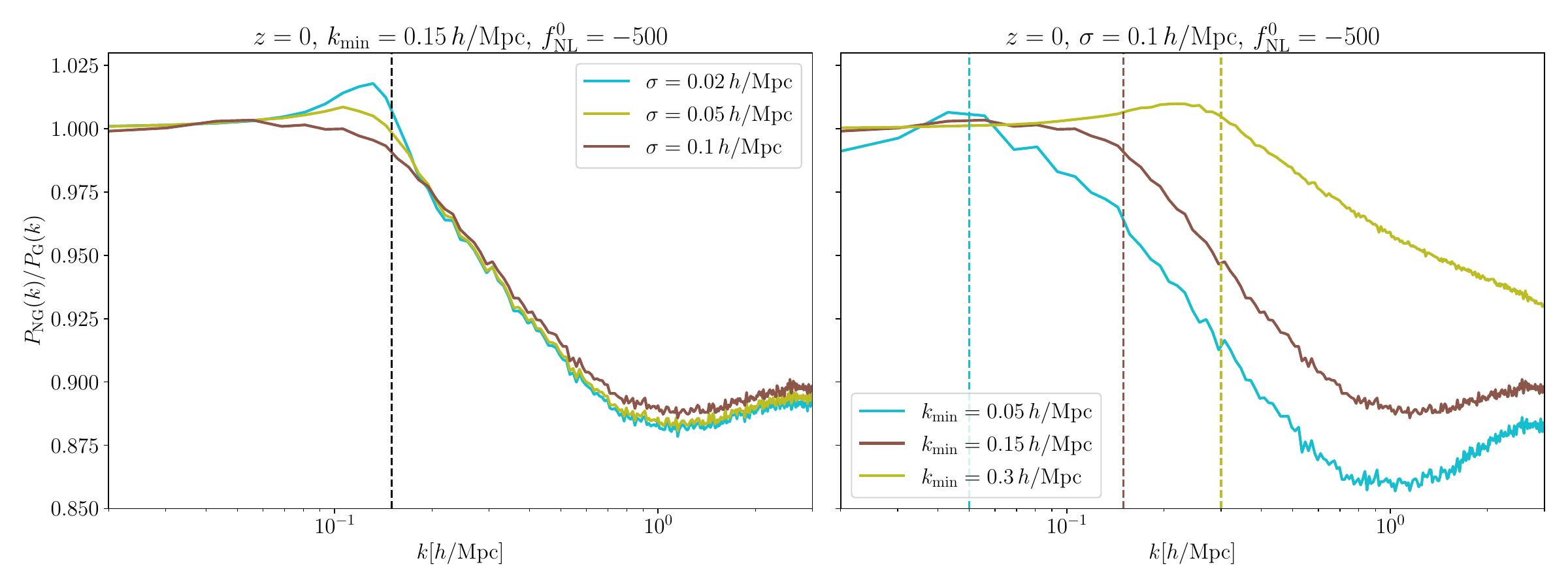}
     \caption{\justifying Ratio of power spectra at $z=0$, illustrating the impact of varying $\sigma$ and $k_{\rm min}$ in Eq.~\eqref{eq:fNLk} to these ratios. On the right panel, we see that $k_{\rm min}$ imposes the scales at which the power spectrum is suppressed, therefore varying it can be an important aspect in alleviating the $S_8$ tension.} 
     \label{fig:impactPM}
 \end{figure*}

\hspace{1cm}
 \section{Halo Mass Function and Void Size Function}
\label{app:HMF}
\begin{figure*}
     \centering
\includegraphics[width=\textwidth]{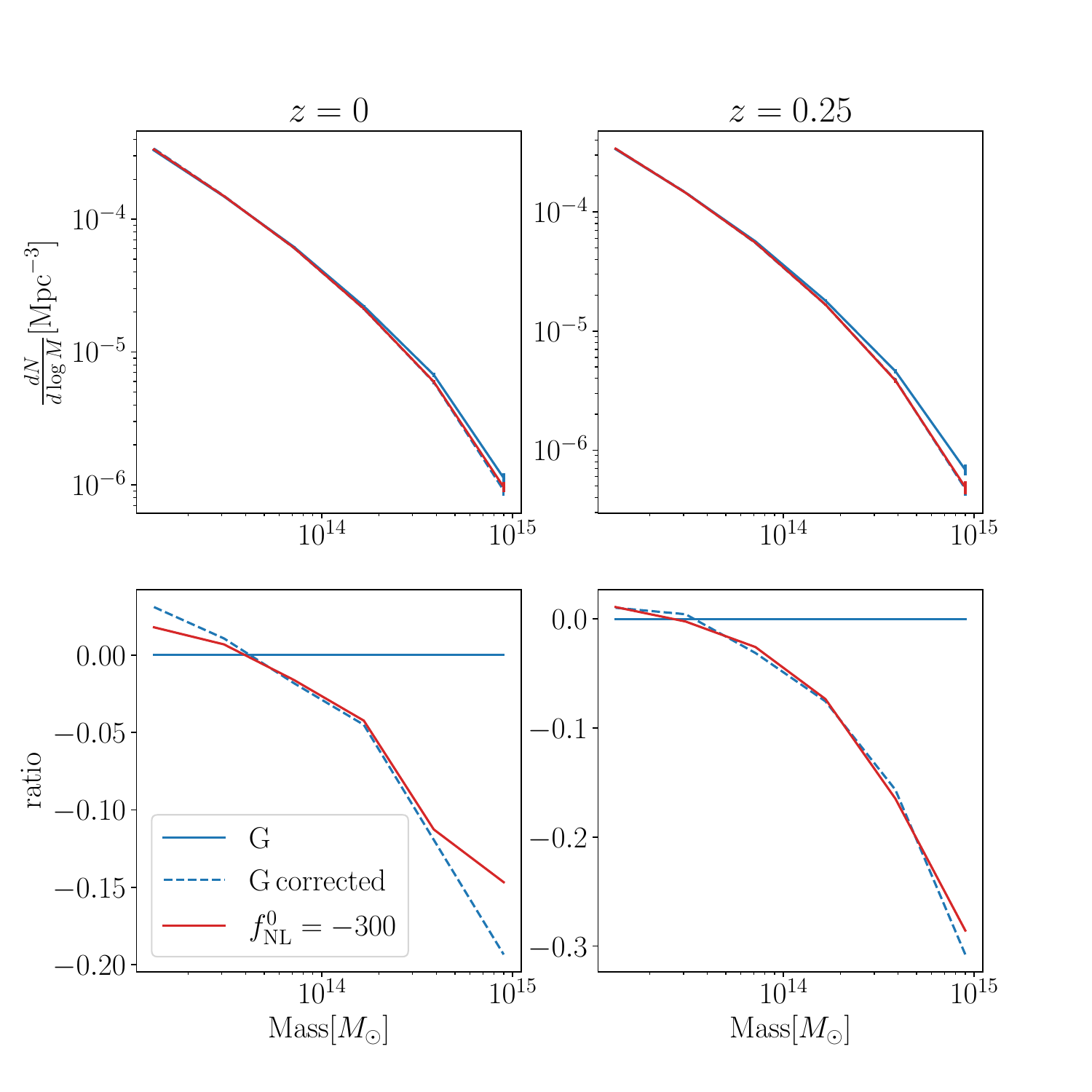}
     \caption{\justifying Halo Mass Function at $z=0$ (left) and $z=0.25$ (right) for the Gaussian and the $f_{\rm NL}^0=-300$ model considered in this work. The bottom panel is the ratio of the HMF with respect to the Gaussian case. Correcting the Gaussian model following Eq.~\eqref{eq:corrM} leads to an effect similar to the non-Gaussian effect, as shown on the bottom panels.}
     \label{fig:HMF}
 \end{figure*}

For a traditional scale-independent global value of $f_{\rm NL}$, local-type PNG with negative $f_{\rm NL}$ are known to induce a drop of the HMF at large mass \cite{Dalal:2007cu,Pillepich:2008ka}. Here, to check how such large masses are affected in our scale-dependent case, we present the HMF for $f_{\rm NL}^0=0$ and $-300$ in Figure \ref{fig:HMF}.

\counterwithin{figure}{section}

It appears that the drop of the HMF would go in the right direction to solve the $S_8$ tension at the cluster count level. To show this a bit quantitatively, we check how the individual mass $M$ in the Gaussian case should vary to reproduce the non-Gaussian case, with the following parametrization \cite{Salvati:2019zdp}:
 \begin{equation}
 \label{eq:corrM}
     M_{\rm corr}=M \cdot {\Upsilon} \cdot \left(\frac{M}{M_*} \right)^{\alpha} \cdot \left( \frac{1+z}{1+z_*} \right)^{\beta},
 \end{equation}
where $M_{*}=2\times 10^{14} M_{\odot}$, $z_*=0.25$ and ($\alpha,\beta,{\Upsilon}$) are allowed to vary. We have found that fixing them to $\alpha=-0.027$, $\beta=0.05$ and ${\Upsilon}=0.92$ roughly matches the Gaussian HMF to the non-Gaussian one, as can be observed in Figure \ref{fig:HMF}. Our results are not mutually exclusive with hydrostatic bias, but could reduce the need for a high bias to reconcile cluster counts with the {\it Planck} value of $S_8$. 

\begin{figure}
     \centering
     \includegraphics[width=.5\textwidth]{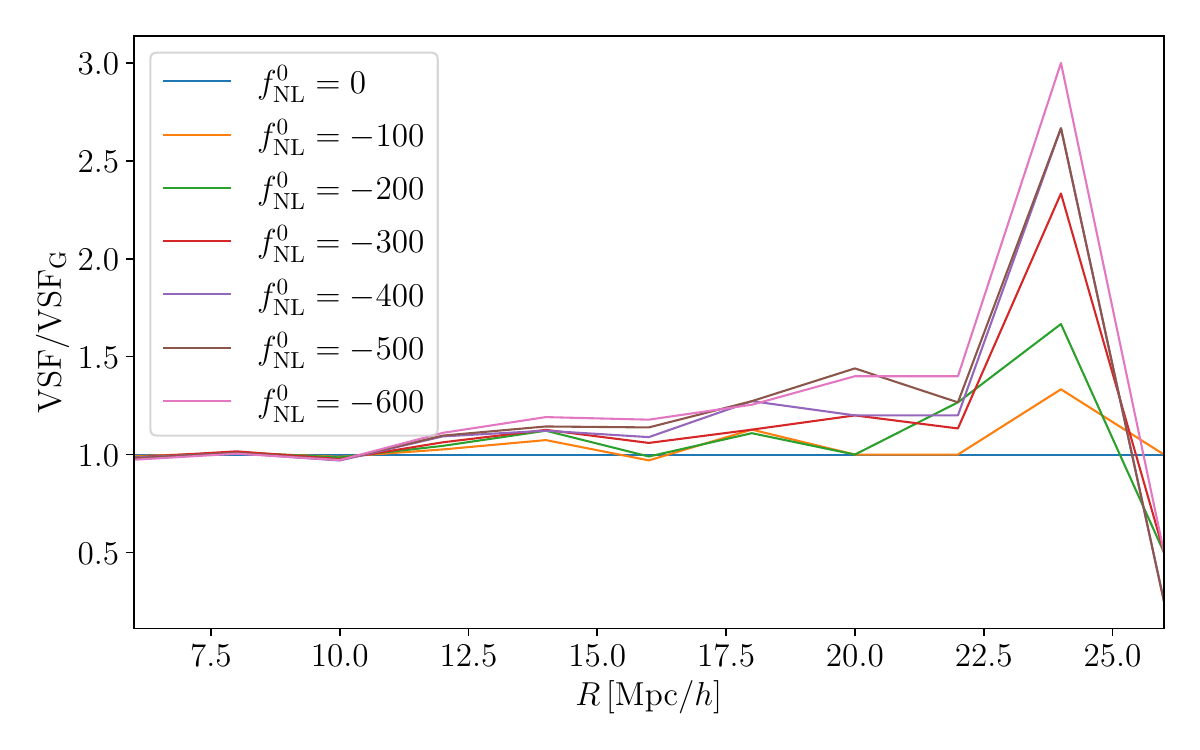}
     \caption{\justifying Ratio of the Void Size Functions (VSF) for the Gaussian and non-Gaussian models considered in this work. The non-Gaussian simulations lead to a larger number of large voids.} 
     \label{fig:VSF}
 \end{figure}

In the same vein, we also studied in our simulations the abundance of underdense regions identified with the void finder of \texttt{Pylians} using an underdensity threshold of $\delta_{\rm t}=-0.7$. The abundance of voids in our simulations, plotted in Figure \ref{fig:VSF}, shows that the largest ($\sim 20$ Mpc$/h$) voids in the $f_{\rm NL}^0=-300$ simulation are twice as numerous at redshift zero than in the Gaussian simulation. This higher proportion of voids may have interesting consequences in the context of the $H_0$ tension \cite{Wong:2021fvu,Contarini:2022nvd,Mazurenko:2023sex}, although the voids that are considered as potentially significant in such a context are an order of magnitude larger, and would certainly need additional modifications to structure formation beyond PNG to explain away the Hubble tension. 

\bibliography{ref}

\end{document}